\documentclass[10pt,twocolumn,english]{revtex4-1}
\usepackage[T1]{fontenc}
\usepackage[latin9]{inputenc}
\usepackage{amsmath}
\usepackage{graphicx}

\makeatletter
\usepackage{babel}
\usepackage{physics}
\usepackage{appendix}

\makeatother

\usepackage{babel}
\begin{document}
\title{Non-local and local temporal cavity soliton interaction in delay models
of mode-locked lasers}
\author{Andrei G. Vladimirov}
\affiliation{Weierstrass Institute, Mohrenstr. 39, 10117 Berlin, Germany}
\begin{abstract}
Interaction equations governing slow time evolution of the coordinates
and phases of two interacting temporal cavity solitons in a delay
differential equation model of a nonlinear mirror mode-locked laser
are derived and analyzed. It is shown that non-local pulse interaction
due to gain depletion and recovery can lead either to a development
of harmonic mode-locking regime, or to a formation of closely packed
incoherent soliton bound state with weakly oscillating intersoliton
time separation. Local interaction via electric field tails can result
in an anti-phase or in-phase stationary and breathing harmonic mode-locking
regimes. 
\end{abstract}
\maketitle
Temporal cavity solitons (TCSs) are short pulses of light circulating
in optical resonators. These solitons were detected experimentally
in driven fiber cavities \cite{leo2010temporal} and optical microcavities
for frequency comb generation \cite{herr2014temporal,kippenberg2018dissipative}.
Another optical system that can support TCSs is a mode-locked (ML)
laser (see \cite{grelu2012dissipative} and references therein) used
for short optical pulse generation. Only those pulses, however, that
are localized in time on a scale much smaller than the cavity round
trip time, can be interpreted as TCSs. In particular, in monolithic
ML semiconductor lasers, where the gain relaxation time is larger
than the cavity round trip time, self-starting ML pulses have a long
gain recovery tail and therefore cannot be considered as localized.
On the contrary, when the cavity round trip time is sufficiently large
as compared to the gain relaxation time the pulses emitted by a passively
ML external-cavity semiconductor laser can be transformed into TCSs
\cite{Marconi}. Note, however, that cavity solitons can also appear
in Haus master equation models of ML fiber lasers where gain is so
slow that it can be considered as constant within the cavity round
trip time, see e. g. \cite{soto1999multisoliton,grelu2004multisoliton,nizette2021generalized}.

Being well separated from one another TCSs can interact via their
exponentially decaying tails. Their interaction in ML lasers was studied
in a number of publications using experimental, numerical and combined
analytical and numerical tools \cite{kutz1998stabilized,akhmediev2001interaction,liu2011interaction,peng2019breathing,ablowitz2009soliton,wang2019optical,grelu2012dissipative,soto1999multisoliton,kokhanovskiy2020single,puzyrev2017bound}.
Many of these studies were performed within the framework of the mean-field
Haus master equations for the case where the gain was either adiabatically
eliminated or constant in time. The effect of the gain (and/or absorption)
saturation and recovery on the pulse interaction in ML lasers was
investigated in \cite{kutz1998stabilized,soto1999multisoliton,nizette2006pulse,ablowitz2009soliton,zaviyalov2012impact,javaloyes2016dynamics,camelin2016electrical,vladimirov2019dynamics}.
In particular, it was shown using a phenomenological approach that
the gain depletion and subsequent recovery can result in repulsive
pulse interaction leading to formation of harmonic mode-locking (HML)
with equally spaced pulses \cite{kutz1998stabilized}. A similar result
was obtained in \cite{nizette2006pulse,camelin2016electrical} using
the delay differential (DDE) model of a passively ML laser developed
in \cite{VT05,VTK,VT04}.

Here using an approach different from that of Ref. \cite{nizette2006pulse}
the interaction of TCSs is studied in a DDE model of nonlinear optical
loop mirror - nonlinear amplifying loop mirror (NOLM-NALM) ML laser
with arbitrary gain relaxation time proposed in \cite{vladimirov2021delay}
(see also Ref. \cite{vladimirov2019dynamics} for the model with adiabatically
eliminated gain). Unlike Haus master equations, DDE models of ML lasers
do not assume small gain and loss per cavity round trip, which means
that ML pulses in these models are always asymmetric. Depending on
the ratio of the spectral filtering width and the gain relaxation
rate two types of TCS interaction are considered, which are referred
below as non-local and local TCS interaction. For both these types
of interaction the equations governing the slow evolution of the time
coordinates and phases of the interacting solitons are derived. It
is shown that apart from the repulsion non-local TCS interaction due
to gain depletion and recovery an attractive interaction is also possible,
which can lead to a closely packed ``incoherent'' oscillating TCS
bound state. Local TCS interaction via electric field component can
lead to a formation of HML regimes with fixed phase difference between
the interacting TCSs as well as breathing HML regimes. The analytical
results obtained here can be used to study the TCS interaction in
the model of passively ML laser \cite{VT05} and other DDE ML laser
models.

A NOLM-NALM ML laser also known as figure-of-eight laser contains
a main cavity with gain medium coupled to a bidirectional nonlinear
mirror loop with intensity dependent reflectivity. Let us consider
the DDE laser model of a NOLM-NALM laser developed in \cite{vladimirov2021delay}
using the lumped element approach described in \cite{VT04,VTK,VT05}:

\begin{gather}
\partial_{t}A+\left(\Gamma+i\omega\right)A=\Gamma\sqrt{\kappa}e^{\frac{1-i\alpha}{2}g_{\tau}+i\theta-i\omega\tau}{\cal R}\left(\left|A_{\tau}\right|^{2}\right)A_{\tau},\label{eq:Model1}\\
\gamma^{-1}\partial_{t}g=p-g-\left(e^{g}-1\right)|A|^{2}\left|{\cal R}\left(\left|A\right|^{2}\right)\right|^{2}.\label{eq:Model2}
\end{gather}
Here $A(t)$ is the electric field envelope, $g(t)$ is the cumulative
gain, $\Gamma$ is the normalized bandwidth of the spectral filter,
$\kappa$ is the linear round-trip attenuation factor, $\alpha$ is
the linewidth extension factor, $\gamma$ is the normalized gain relaxation
rate, $p$ is the pump parameter, $\theta$ is a phase shift, and
$\omega$ is the reference frequency, which will be chosen below.
The subscript $\tau$ denotes delayed argument, where the delay time
$\tau$ is equal to the dimensionless cold cavity round trip time.
Complex reflectivity of the nonlinear mirror is defined by the relation:
\[
{\cal R}\left(\left|A\right|^{2}\right)=\sqrt{G}\left[\left(1-K\right)e^{-i\chi\left(1-K\right)\left|A\right|^{2}}-Ke^{-i\chi KG\left|A\right|^{2}}\right],
\]
where $0<K<1$ is the beam splitting ratio, $G<1$ ($G>1$) is the
linear attenuation (amplification) in the nonlinear mirror loop, and
$\chi$ is the normalized Kerr coefficient. In particular, $K=0.5$
corresponds to a symmetric splitter. The quantity $\left|{\cal R}\right|^{2}$
in Eq. (\ref{eq:Model2}) is the intensity reflectivity of the nonlinear
mirror \cite{doran1988nonlinear,lai2005nolm,fermann1990nonlinear}.
This quantity oscillates with the intensity between the minimal value
$\left|{\cal R}\right|^{2}=G\left(1-2K\right)^{2}$ and the maximal
value $\left|{\cal R}\right|^{2}=G$, where the first minimum is achieved
at $\left|A\right|^{2}=0$ and corresponds to zero reflectivity $\left|{\cal R}\right|^{2}=0$
in the case of symmetric splitter, $K=0.5$. Note that the model equations
(\ref{eq:Model1}) and (\ref{eq:Model2}) neglect the chromatic dispersion
of the intracavity media. An approach to account for the material
dispersion in DDE laser models was proposed in \cite{pimenovprl,pimenov2020temporal}.

\begin{figure}
\includegraphics[scale=0.33]{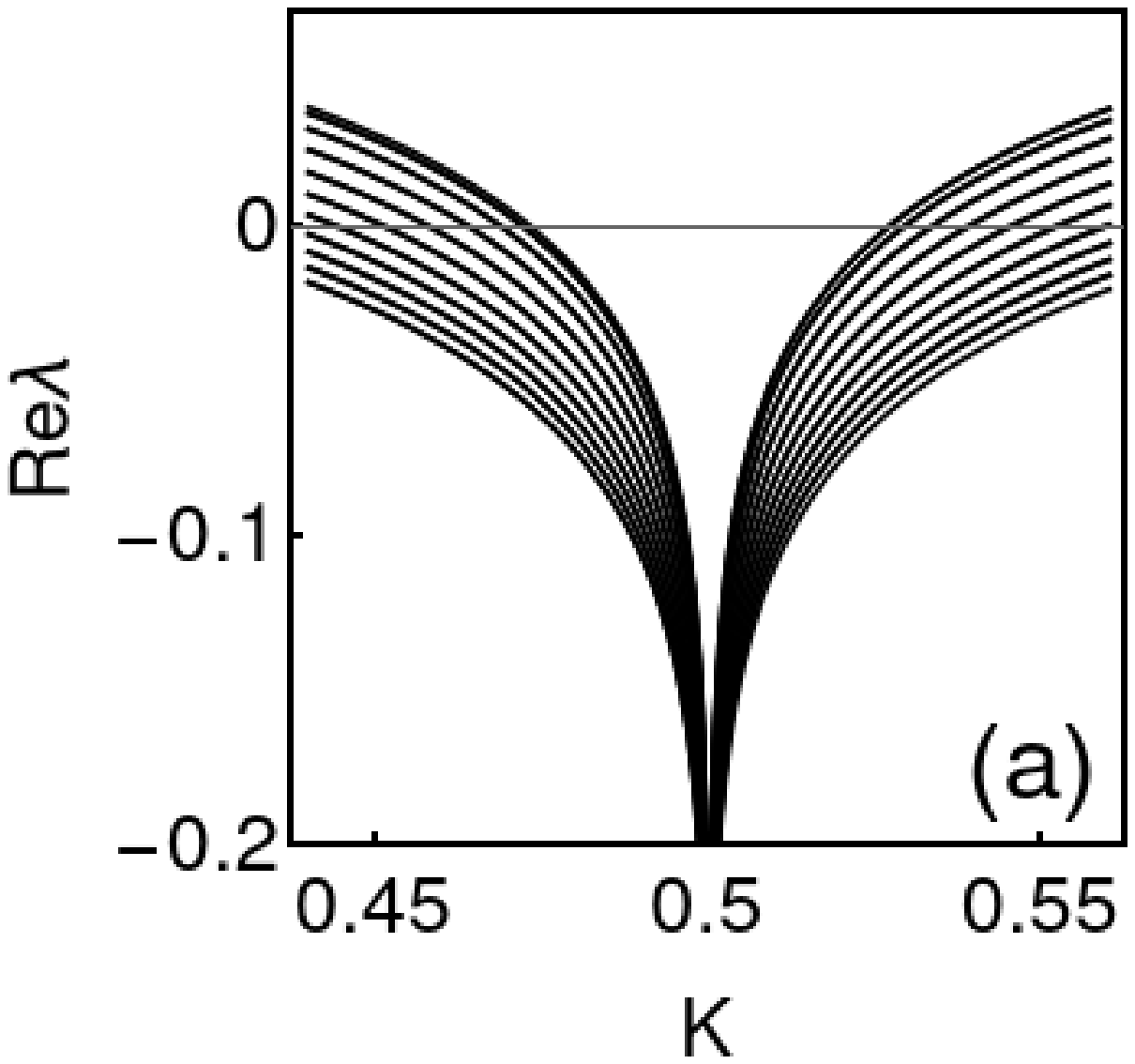}\includegraphics[scale=0.33]{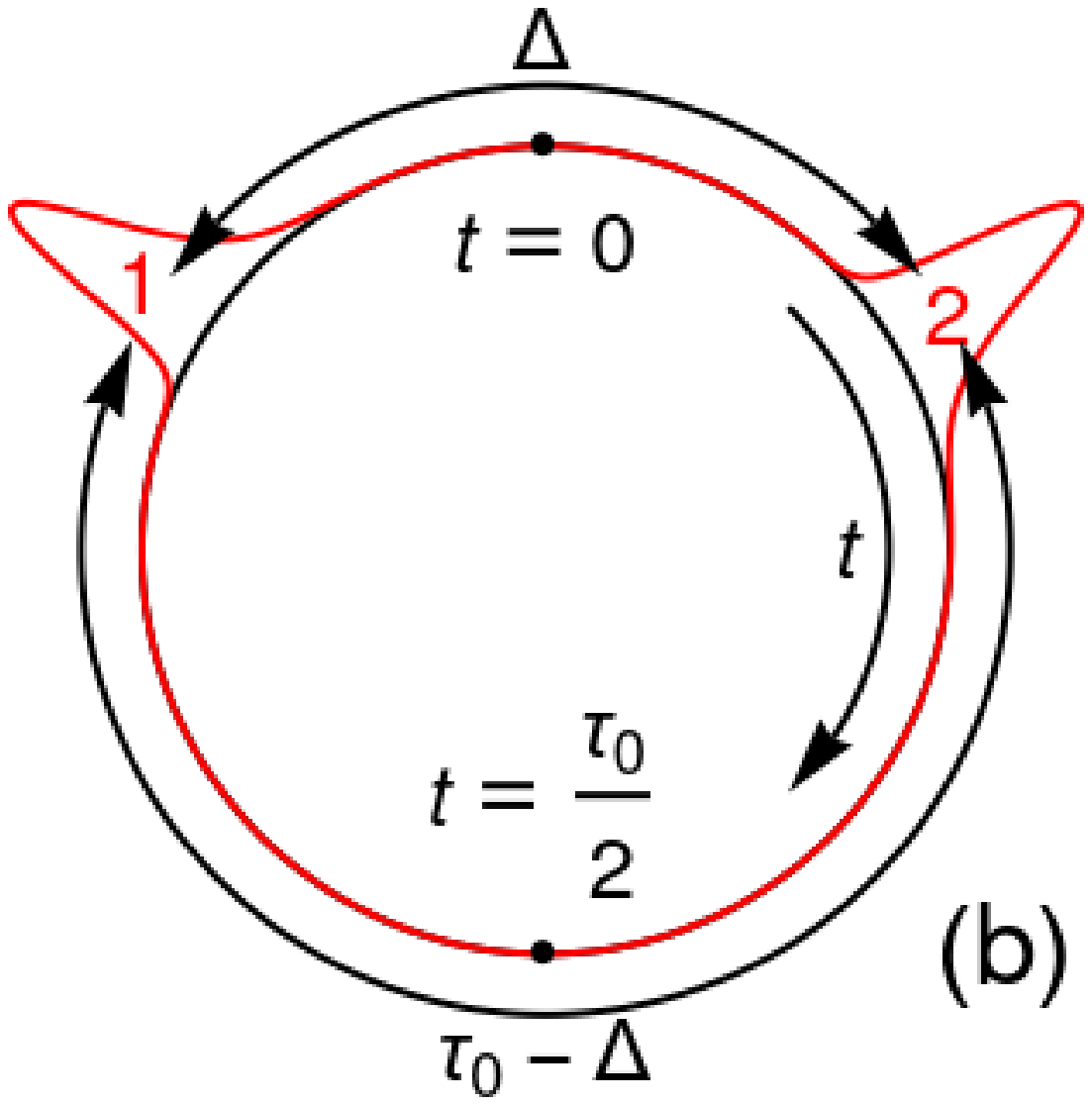}

\caption{Largest real parts of the eigenvalues $\lambda$ defined by Eq. (\ref{eq:Lambert})
(a). Schematic representation of two interacting TCSs labeled as 1
and 2 (b). $\Delta$ is the TCS time separation and $\tau_{0}$ is
the TCS period.\label{fig:scheme}}
\end{figure}

Linear stability of the trivial solution of Eqs. (\ref{eq:Model1})
and (\ref{eq:Model2}), $A=0$ and $g=p$, is determined by the equation
\begin{equation}
\partial_{t}A+\left(\Gamma+i\omega\right)A=\Gamma\sqrt{\kappa}e^{\frac{1-i\alpha}{2}p+i\theta-i\omega\tau}{\cal R}\left(0\right)A_{\tau},\label{eq:lin}
\end{equation}
giving an infinite set of eigenvalues that can be expressed in terms
of the Lambert function:
\begin{equation}
\lambda_{k}=-\Gamma-i\omega+\frac{1}{\tau}W_{k}\left[\Gamma\tau\sqrt{\kappa}{\cal R}\left(0\right)e^{\Gamma\tau+\frac{1-i\alpha}{2}p+i\theta}\right]\label{eq:Lambert}
\end{equation}
with $k=0,\pm1\pm2\dots$. Real parts of these eigenvalues for $-10\le k\le10$
are shown in Fig. \ref{fig:scheme}(a) as functions of the splitting
ratio $K$. It is seen that the trivial solution is stable when $K$
is sufficiently close to $0.5$. In particular, the real parts of
all the eigenvalues tend to $-\infty$ for $K\to0.5$, except for
a single eigenvalue $\lambda_{0}\to-\left(\Gamma+i\omega\right)$.
Therefore, the use of almost symmetric beam splitter is needed to
achieve the TCS regime in Eqs. (\ref{eq:Model1}) and (\ref{eq:Model2}). 

In order to derive the interaction equations governing the slow evolution
of the time coordinates and phases of two well separated interacting
TCSs let us first rewrite the model equations (\ref{eq:Model1}) and
(\ref{eq:Model2}) in a more general form:

\begin{equation}
\partial_{t}{\bf U}={\bf F}\left({\bf U},\omega\right)+{\bf H}\left({\bf U}_{\tau},\omega\right),\label{eq:general_form}
\end{equation}
where the column vector ${\bf U}=\left(\begin{array}{ccc}
\Re A & \Im A & z\end{array}\right)^{T}$ , $z=g-p$, ${\bf F}\left(0,\omega\right)={\bf H}\left(0,\omega\right)=0$,
and subscript $\tau$ denotes time delay. Furthermore, in our case
similarly to the DDE ML laser model in \cite{VT05} only two first
components of the vector ${\bf H}$ are nonzero, see Eqs. (\ref{eq:Model1})
and (\ref{eq:Model2}).

Let us assume that the cavity round trip is sufficiently large, $\tau\gg\Gamma^{-1},\gamma^{-1}$
and consider $\tau_{0}$-periodic TCS solution of Eq. (\ref{eq:general_form})
defined by ${\bf U={\bf u}}_{0}$ and $\omega=\omega_{0}$, where
${\bf u}_{0}\left(t\right)={\bf u}_{0}\left(t+\tau_{0}\right)$ with
${\bf u}_{0}=\left(\begin{array}{ccc}
\Re A_{0} & \Im A_{0} & z_{0}\end{array}\right)^{T}$, the period $\tau_{0}$ is close to the delay time $\tau$.

The decay rates of the TCS tails are determined by the linear equation
(\ref{eq:lin}), where the delay time $\tau$ and the frequency offset
$\omega$ are replaced with $-\delta=\tau-\tau_{0}$ and $\omega_{0}$,
respectively \cite{vladimirov2019dynamics,yanchuk2019temporal}, and
the linearization of Eq. (\ref{eq:Model2}) on the trivial solution:
\begin{equation}
\partial_{t}A+\left(\Gamma+i\omega_{0}\right)A=\Gamma\sqrt{\kappa}e^{\frac{1-i\alpha}{2}p+i\theta+i\omega_{0}\delta}{\cal R}\left(0\right)A_{-\delta},\label{eq:LinearA}
\end{equation}
\begin{equation}
\partial_{t}z=-\gamma z.\label{eq:LinearG}
\end{equation}
Equation (\ref{eq:LinearG}) assumes that the decay rate of the TCS
gain component at $t\to+\infty$ is determined by the eigenvalue $-\gamma$,
while Eq. (\ref{eq:LinearA}) has an infinite number of eigenvalues
defined by Eq. (\ref{eq:Lambert}), where $\tau$ and $\omega$ are
replaced with $-\delta$ and $\omega_{0},$ respectively. However,
in the limit $K\to0.5$, where ${\cal R}\left(0\right)\to0$, only
a single linear eigenvalue $-\left(\Gamma+i\omega_{0}\right)$ with
negative real part remains, while the real parts of all the other
eigenvalues tend to $+\infty$. Hence, in this limit the leading edge
of the TCS decays faster than exponentially \cite{vladimirov2019dynamics},
whereas the field component of the trailing edge of the TCS contains
only a single decaying exponent. Therefore, in a laser with symmetric
beam splitter the TCS asymptotical behavior at sufficiently large
positive times $t>0$ is described by: 
\begin{equation}
A_{0}\sim ae^{-\left(\Gamma+i\omega_{0}\right)t},\quad z_{0}\sim be^{-\gamma t}.\label{eq:TCS}
\end{equation}
Here $a$($b$) is complex (real) coefficient, which depends on the
particular form of the pulse solution and can be calculated numerically.
Since the leading edge of the pulse decays faster than exponentially
it can be neglected when constructing the interaction equations. Note,
that in Eq. (\ref{eq:TCS}) it is assumed that the pulse is positioned
at the origin of the coordinate $t$. In the case where $\Gamma>\gamma$,
the gain tail behind the TCS is longer than the tail of the electromagnetic
field and two well separated TCSs interact via their gain components.
Below this type of interaction will be referred to as ``non-local''
interaction. When, on the other hand, $\Gamma<\gamma$ the gain tail
is shorter than the field tail and the TCSs interact via the field
components. This second type of interaction will be further referred
to as ``local'' interaction. Local interaction can take place, in
particular, when the gain variable is eliminated adiabatically \cite{vladimirov2019dynamics}.
Note, that a small asymmetry of the splitter does not change the analytical
results presented below and only slightly modifies the decay rate
of the TCS trailing tail. Sufficiently large asymmetry on the other
hand can destabilize the trivial solution \cite{vladimirov2021delay}
thus breaking a necessary condition of the TCS formation.

Linear stability of the TCS solution ${\bf U=u}_{0}$ and $\omega=\omega_{0}$
is determined by linearizing Eq. (\ref{eq:general_form}) on this
solution and calculating the spectrum of the resulting linear operator
${\cal L}$. Due to the translational and phase shift symmetries of
the model equations, $A(t)\to A(t-t_{0})$ and $A(t)\to A(t)e^{i\phi_{0}}$
with arbitrary constant $t_{0}$ and $\phi_{0}$, the operator ${\cal L}$
has a double zero eigenvalue, ${\cal L}{\bf v}^{(\tau,\phi)}=0$,
where the translational and phase shift neutral (Goldstone) modes
are ${\bf v}^{(\tau)}=\partial_{t}{\bf u}_{0}^{(\tau)}$ and ${\bf v}^{(\phi)}=\left(\begin{array}{ccc}
-\Im A_{0} & \Re A_{0} & 0\end{array}\right)^{T}$, respectively. Let us assume that the TCS is stable, which means
that the rest of the spectrum of ${\cal L}$ lies in the left half
of the complex plane. Similarly, the adjoint lnear operator ${\cal L}^{\dagger}$
has a double zero eigenvalue associated with the so-called adjoint
neutral modes ${\bf w}^{(\tau,\phi)}$, ${\cal L}^{\dagger}{\bf w}^{(\tau,\phi)}=0$.
Since the adjoint operator ${\cal L}^{\dagger}$ is obtained from
${\cal L}$ by the transformations including the time reversal, $t\to-t$
(see Appendix), the asymptotic behavior of the adjoint neutral modes
at sufficiently large negative times $t<0$ is given by 
\begin{equation}
\xi^{(\tau,\phi)}\sim c^{(\tau,\phi)}e^{\left(\Gamma-i\omega_{0}\right)t},\quad\zeta_{g}^{(\tau,\phi)}\sim d^{(\tau,\phi)}e^{\gamma t},\label{eq:adjoint_asympt}
\end{equation}
where the adjoint neural mode is given by the row vector ${\bf w}^{(\tau,\phi)}=\left(\begin{array}{ccc}
\Re\xi^{(\tau,\phi)} & \Im\xi^{(\tau,\phi)} & \zeta_{g}^{(\tau,\phi)}\end{array}\right)$ and $c^{(\tau,\phi)}$ ($d^{(\tau,\phi)}$) are complex (real) coefficients,
which can be calculated numerically. Similarly to the leading tail
of the TCS solution, the trailing tail of the adjoint neutral modes
decay faster than exponentially at large $t>0$. The adjoint neutral
modes are assumed to satisfy the bi-orthogonality condition $\int_{0}^{\tau_{0}}{\bf w}^{(j)}\cdot{\bf v}^{(k)}dt=\delta_{jk}$
with $j,k=\tau,\phi$.

Let us look for the solution of Eq. (\ref{eq:general_form}) in the
form of a sum of two well separated TCSs 
\begin{eqnarray}
{\bf U} & = & {\bf u}_{\Sigma}+\delta{\bf u},\quad{\bf u}_{\Sigma}={\bf u}_{1}+{\bf u}_{2},\label{eq:Anzatz}
\end{eqnarray}
where ${\bf u}_{1,2}=\left(\begin{array}{ccc}
\Re A_{1,2} & \Im A_{1,2} & g_{1,2}\end{array}\right)^{T}$ with $A_{1,2}=A_{0}\left(t-\tau_{1,2}\right)e^{i\phi_{1,2}}$ and
$g_{1,2}=z_{0}\left(t-\tau_{1,2}\right)$, plus a small correction
$\delta{\bf u={\cal O}}\left(\epsilon\right)$. Here the small parameter
$\epsilon$ characterizes the weak overlap of the TCSs. Coordinates
$\tau_{1,2}$ and phases $\phi_{1,2}$ of the interacting TCSs are
assumed to be slow functions of time, $\partial_{t}\tau_{1,2},\partial_{t}\phi_{1,2}={\cal O}\left(\epsilon\right)$.

Substituting Eq. (\ref{eq:Anzatz}) into Eq. (\ref{eq:general_form}),
collecting the first order terms in small parameter $\epsilon$, and
using the solvability conditions \cite{halanay1966differential} yields
\begin{gather}
\partial_{t}\tau_{1,2}=-\left\langle {\bf w}_{1,2}^{(\tau)}{\bf P}\right\rangle ,\quad\partial_{t}\phi_{1,2}=\left\langle {\bf w}_{1,2}^{(\phi)}{\bf P}\right\rangle ,\label{eq:Int1_tau_phi}\\
{\bf P=}-\partial_{t}\mathbf{u}_{\Sigma}+{\bf F}\left({\bf u}_{\Sigma},\omega_{0}\right)+{\bf G}\left[{\bf u}_{\Sigma}\left(t-\tau\right),\omega_{0}\right],\label{eq:P}
\end{gather}
where $\left\langle \cdot\right\rangle =\int_{0}^{\tau_{0}}\cdot dt$,
${\bf w}_{1,2}^{(\tau,\phi)}=\left(\begin{array}{ccc}
\Re\xi_{1,2}^{(\tau,\phi)} & \Im\xi_{1,2}^{(\tau,\phi)} & \zeta_{1,2}^{(\tau,\phi)}\end{array}\right)$ with $\xi_{1,2}^{(\tau,\phi)}=\xi^{(\tau,\phi)}\left(t-\tau_{1,2}\right)e^{-i\phi_{1,2}}$
and $\zeta_{1,2}^{(\tau,\phi)}=\zeta^{(\tau,\phi)}\left(t-\tau_{1,2}\right)$.
Similarly to the case of dissipative soliton interaction in partial
differential equation models, the right hand side (RHS) of the interaction
equations (\ref{eq:Int1_tau_phi}) can be expressed in terms of the
TCS solution itself and the adjoint neutral modes, see, e.g. \cite{vladimirov2001stable,vladimirov2018effect,vladimirov2021dissipative}.
The details of the calculations are given in the Appendix. As a result,
the interaction equations for $\tau_{0}$-periodic TCS take the form
\begin{gather}
\partial_{t}\tau_{1,2}=\pm{\bf w}_{1,2}^{(\tau)}\left(0\right){\bf u}_{2,1}\left(0\right)\mp{\bf w}_{1,2}^{(\tau)}\left(\frac{\tau_{0}}{2}\right){\bf u}_{2,1}\left(\frac{\tau_{0}}{2}\right),\label{eq:int2_tau}\\
\partial_{t}\phi_{1,2}=\mp{\bf w}_{1,2}^{(\phi)}\left(0\right){\bf u}_{2,1}\left(0\right)\pm{\bf w}_{1,2}^{(\phi)}\left(\frac{\tau_{0}}{2}\right){\bf u}_{2,1}\left(\frac{\tau_{0}}{2}\right),\label{eq:int2_phi}
\end{gather}
where without the loss of generality one can assume that $t=0$ and
$t=\tau_{0}/2$ correspond respectively to the middle point between
two TCS and the opposite point on a circle with the circumference
$\tau_{0}$, see Fig. \ref{fig:scheme}. Note that due to the super-exponential
decay of the TCS leading and adjoint neutral modes trailing tails
the terms ${\bf w}_{1}^{(\tau,\phi)}\left(0\right){\bf u}_{2}\left(0\right)$
and ${\bf w}_{2}^{(\tau,\phi)}\left(\tau_{0}/2\right){\bf u}_{1}\left(\tau_{0}/2\right)$
in Eqs. (\ref{eq:int2_tau}) and (\ref{eq:int2_phi}) can be neglected.
The remaining terms ${\bf w}_{2}^{(\tau,\phi)}\left(0\right){\bf u}_{1}\left(0\right)$
and ${\bf w}_{1}^{(\tau,\phi)}\left(\tau_{0}/2\right){\bf u}_{2}\left(\tau_{0}/2\right)$
entering the RHS of the equations for $\tau_{2}$($\phi_{2}$) and
$\tau_{1}$($\phi_{1}$) respectively, have very different magnitudes
except for the case where the TCSs are close to equidistant in the
cavity, $\Delta=\tau_{2}-\tau_{1}\approx\tau_{0}/2$. This means that
except for this case the TCS interaction is strongly asymmetric and
does not satisfy third Newton's law \cite{camelin2016electrical,vladimirov2018effect}.

Let us first consider the case of non-local interaction, $\Gamma>\gamma$.
Here the gain component of the trailing tail of the TCS decays slower
than the field component and, hence, TCSs interact via the gain tail
behind the pulse. This interaction will be called non-local interaction
since the gain tail can last much longer than the duration of the
electromagnetic pulse. Therefore, when the time separation of two
TCSs is sufficiently large the terms proportional to $e^{-\Gamma t}$($e^{\Gamma t}$)
in the asymptotic expressions (\ref{eq:TCS}) and (\ref{eq:adjoint_asympt})
for the TCS and adjoint neutral modes can be neglected. Then substituting
these asymptotic expressions into Eqs. (\ref{eq:int2_tau}) and (\ref{eq:int2_phi})
yields 
\begin{gather}
\partial_{t}\Delta=-s^{(\tau)}\left[e^{-\gamma\Delta}-e^{-\gamma(\tau_{0}-\Delta)}\right],\label{eq:nonlocal1}\\
\partial_{t}\psi=s^{(\phi)}\left[e^{-\gamma\Delta}-e^{-\gamma(\tau_{0}-\Delta)}\right],\label{eq:nonlocal2}
\end{gather}
where $\Delta=\tau_{2}-\tau_{1}$ and $\psi=\phi_{2}-\phi_{1}$ are
the TCS time separation and phase difference, $s^{(\tau,\phi)}=bd^{(\tau,\phi)}$.
The interaction equations (\ref{eq:nonlocal1}) and (\ref{eq:nonlocal2})
have a single stationary solution corresponding to a HML regime with
two equidistant pulses in the cavity, $\Delta\tau=\tau_{0}/2$. This
solution is stable for $s^{(\tau)}<0$ and unstable for $s^{(\tau)}>0$
. The first case corresponds to the repulsion of two TCSs on a circle
when the distance between them is increasing until the pulses become
equidistant. In the second case two TCS are attracted to one another.

\begin{figure}
\includegraphics[scale=0.43]{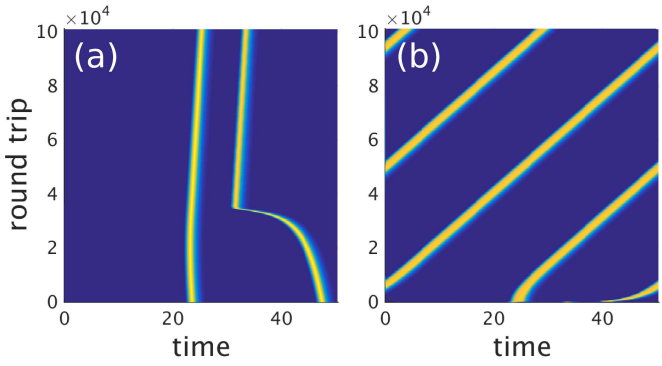} \caption{Interaction of two TCSs leading to a closely packed incoherent TCS
bound state at $p=5.5$ (a) and a stable HML with two equidistant
pulses circulating in the laser cavity at $p=5.7$ (b). Other parameters
are: $T=50.$, $\Gamma=1$, $\kappa=0.8$, $\alpha=0$, $\gamma=0.3$,
$G=0.5$, $\chi=2$.\label{fig:nonlocal}}
\end{figure}

\begin{figure}
\includegraphics[scale=0.35]{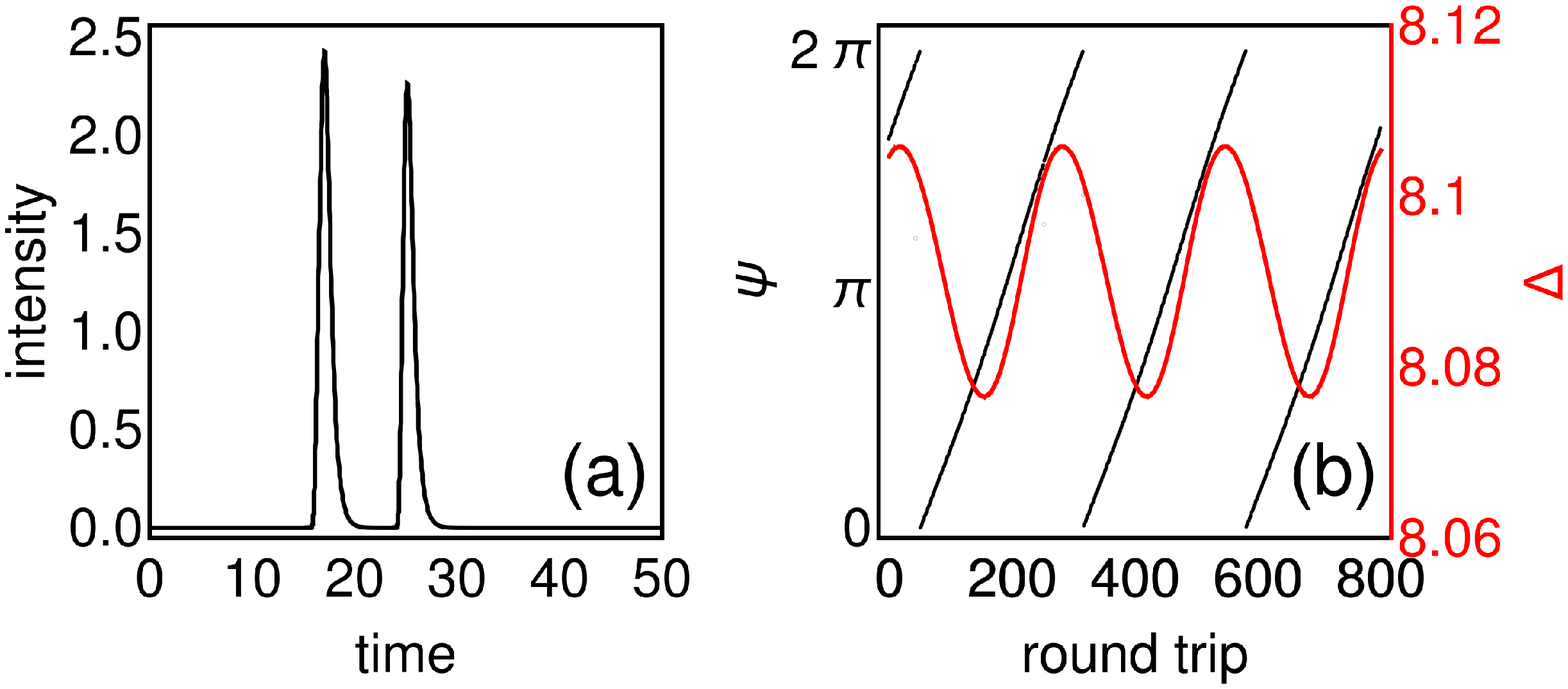}

\caption{Intensity time trace of the closely packed TCS bound state (a). Intersoliton
phase difference $\psi$ and time separation $\Delta$ for this bound
state as a functions of round trip number (b). Parameters are the
same as for Fig. \ref{fig:nonlocal}(a).\label{fig:BS}}
\end{figure}

Figure \ref{fig:nonlocal}(a) illustrates the evolution of the TCS
coordinates with the round trip number calculated by numerical integration
of the model equations (\ref{eq:Model1}) and (\ref{eq:Model2}).
In this figure corresponding to positive $s^{(\tau)}=0.432$ and $s^{(\phi)}=-3.531$
the first TCS attracts the second one until the distance between them
becomes sufficiently small and a closely packed bound state of two
TCS is formed. The intensity time trace of this bound state is shown
in Fig. \ref{fig:BS}(a) where the peak power of the first TCS is
larger than that of the second one. Similar bound states were observed
experimentally in a NOLM-NALM figure-of-eight ML laser in \cite{kokhanovskiy2020single}.
Fig. \ref{fig:BS}(b) obtained by numerical integration of the model
equations (\ref{eq:Model1}) and (\ref{eq:Model2}) illustrates how
the TCS phase difference $\psi$ and time separation $\Delta$ evolve
with the cavity round trip number. It is seen that similarly to the
``type-A'' bound states reported in a Haus model of passively ML
laser with a slow absorber \cite{soto1999multisoliton} the phase
difference $\psi$ grows monotonously in time. Therefore, this bound
state can be called ``incoherent.'' Furthermore, it is seen that
the TCS time separation $\Delta$ is in fact not stationary, but exhibits
small amplitude oscillations with the period equal to the time interval
during which $\psi$ changes by $2\pi$. This indicates that the oscillations
of $\Delta$ are due to local interaction of the TCSs via exponentially
decaying tails of the electromagnetic field. Note that since the dependence
of $\psi$ on time is not strictly linear {[}see Fig. \ref{fig:BS}(b){]},
the time-averaged local interaction force between the two TCSs can
be nonzero. Note, however, that the approach based on the properties
of the unperturbed identical TCS solutions and their adjoint neutral
modes is hardly applicable to describe the bound state shown in Fig.
\ref{fig:BS}(a), where the peak power of the second interacting TCS
is noticeably smaller than that of the first one.

Figure \ref{fig:nonlocal}(b) corresponds to the case of negative
$s^{(\tau)}=-2.643$ leading to TCS repulsion and $s^{(\phi)}=-3.599$.
In this figure the second TCS is repelled from the first one until
the two TCS become equidistant in time, which corresponds to a HML
regime with two pulses per cavity round trip.

\begin{figure}
\includegraphics[scale=0.43]{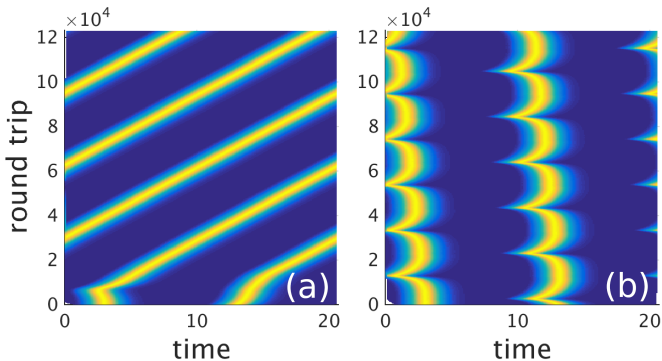}

\caption{Local TCS interaction leading to a development of stationary bistable
anti-phase (a) and breathing (b) HML regimes calculated numerically
using different initial conditions. $p=6.8$, $\tau=20$, $\gamma=10$.
Other parameters are the same as for Fig. \ref{fig:nonlocal}.}
\label{fig:local} 
\end{figure}

\begin{figure}
\includegraphics[scale=0.3]{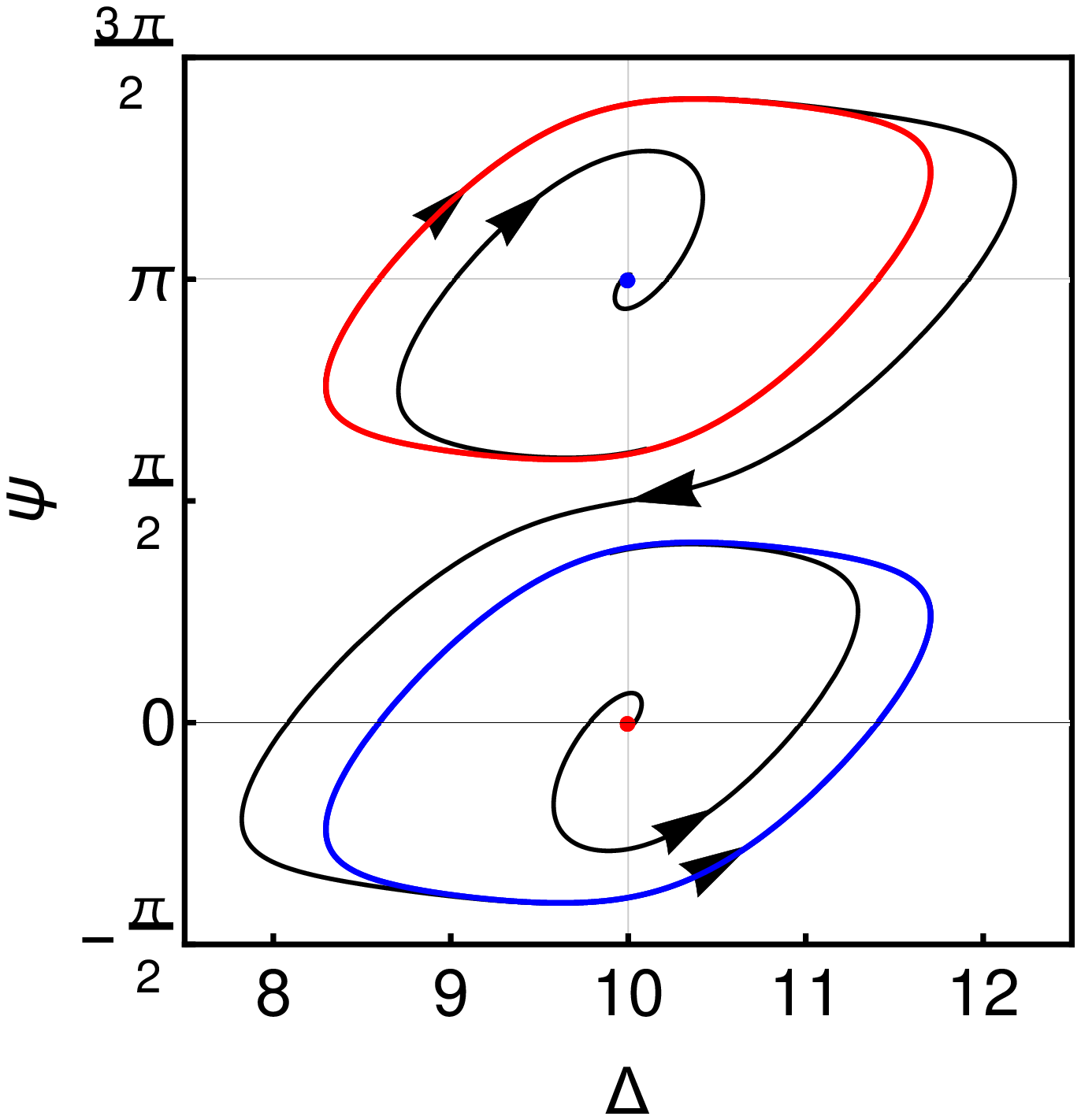}\includegraphics[scale=0.3]{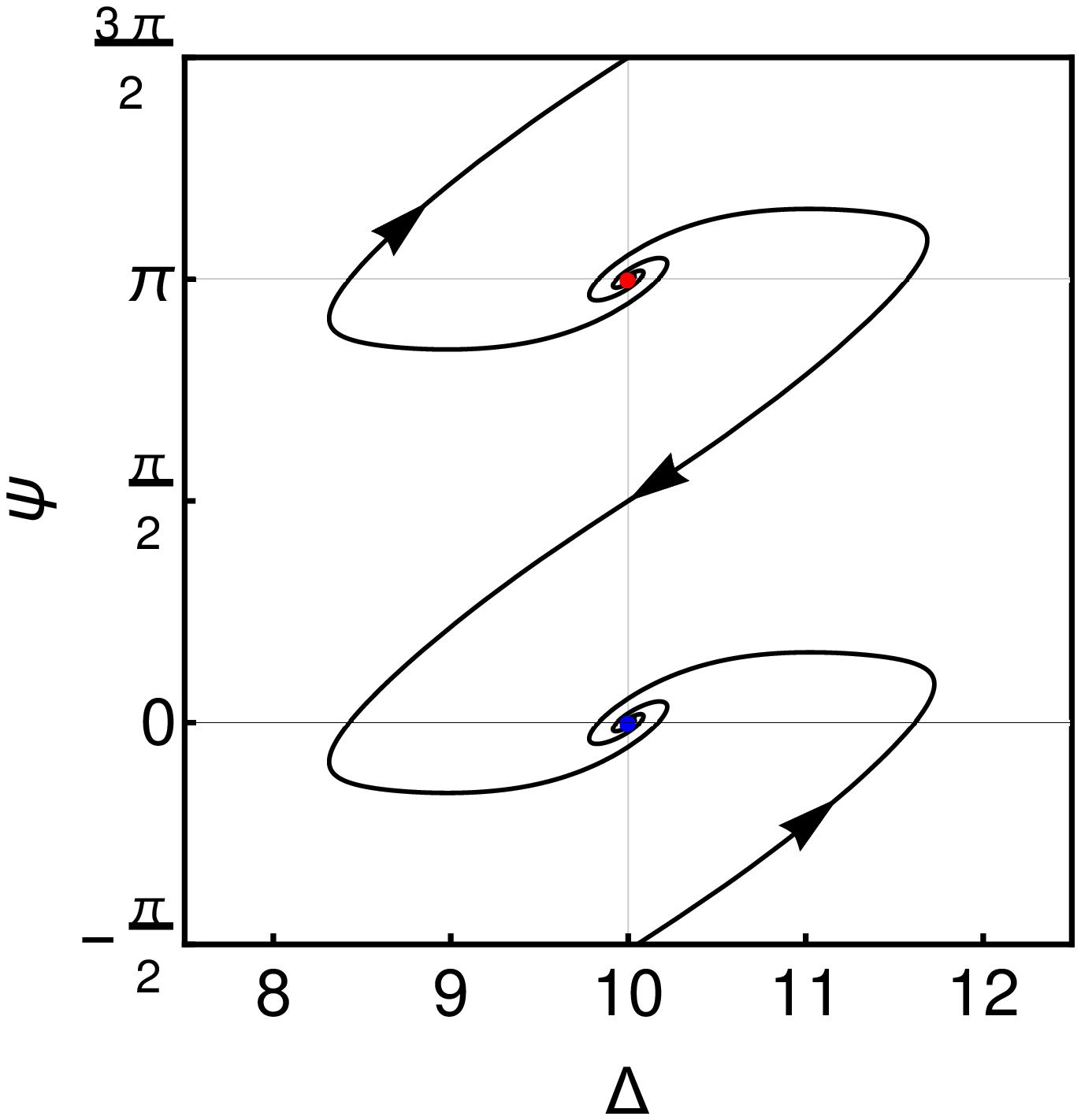}

\caption{Phase plane of Eqs. (\ref{eq:Local1}) and (\ref{eq:Local2}) calculated
for $p=6.8$ (a) and $p=7.2$ (b). Stable (unstable) limit cycle and
steady states are shown by blue (red) color. Other parameters are
the same as in Fig. \ref{fig:local}.\label{fig:phase}}
\end{figure}

In the case where $\Gamma<\gamma$ the gain tail behind the pulse
is shorter than that of the electric field. In this case the field
component dominates the interaction of two well separated TCS and
the interaction equations take the form:
\begin{gather}
\partial_{t}\Delta=-\Re\left[q^{(\tau)}f\left(\Delta,\psi\right)\right],\quad\partial_{t}\psi=\Re\left[q^{(\phi)}f\left(\Delta,\psi\right)\right],\label{eq:Local1}\\
f\left(\Delta,\psi\right)=e^{-\left(\Gamma+i\omega_{0}\right)\Delta-i\psi}-e^{-\left(\Gamma+i\omega_{0}\right)\left(\tau_{0}-\Delta\right)+i\psi},\label{eq:Local2}
\end{gather}
where $q^{(\tau,\phi)}=a\bar{c}^{(\tau,\phi)}$ and bar denotes complex
conjugation. These equations have two steady states with equal pulse
separations, $\Delta=\tau_{0}/2$, and opposite phase differences,
$\psi=0$ and $\psi=\pi$, on the interval $\psi\in\left[0,2\pi\right)$.
They correspond to HML regimes with equidistant in time in-phase and
anti-phase TCSs. Due to the symmetry property of Eqs. (\ref{eq:Local1})
and (\ref{eq:Local2}), $t\to-t$ and $\psi\to\psi+\pi$, if one of
the two steady states is asymptotically stable, another is unstable,
and vice versa. Furthermore, Eqs. (\ref{eq:Local1}) and (\ref{eq:Local2})
can exhibit an Andronov-Hopf bifurcation at $\Re\left\{ \left[q^{(\tau)}\left(\Gamma+i\omega_{0}\right)-iq^{(\phi)}\right]e^{-i\omega_{0}\tau_{0}/2}\right\} =0$,
where a pair of limit cycles of opposite stability (stable and unstable)
bifurcate simultaneously from the in-phase and anti-phase steady states.
These two cycles are shown in Fig. \ref{fig:phase}(a) for the parameter
values of Fig. \ref{fig:local}, which correspond to $q^{(\tau)}=-9.386+10.312i$,
$q^{(\phi)}=5.870-16.291i$, and $\omega_{0}=0.183427$. It is seen
that in agreement with the results of numerical simulations shown
in Fig. \ref{fig:local} two stable attractors, fixed point with $\psi=\pi$
and a stable limit circle born from the fixed point with $\psi=0$,
coexist in the phase plane of the interaction equations. With the
increase of the pump parameter $p$ limit cycles shown in Fig. \ref{fig:phase}(a)
shrink and disappear in the inverse Andronov-Hopf bifurcations of
the corresponding fixed points. This bifurcation stabilizes (destabilizes)
the fixed point with $\psi=0$ ($\psi=\pi$). The resulting phase
portrait of Eqs. (\ref{eq:Local1}) and (\ref{eq:Local2}) with $q^{(\tau)}=-19.633+7.608i$,
$q^{(\phi)}=17.251-11.001i$, and $\omega_{0}=0.181633$ calculated
numerically for $p=7.2$ is shown in Fig. \ref{fig:phase}(b). This
figure is in agreement with the direct numerical simulations of the
model equations (\ref{eq:Model1}) and (\ref{eq:Model2}) which indicate
that a stable HML regime with two in-phase pulses is formed as a result
of the local TCS interaction at $p=7.2$. Note, that in a laser with
sufficiently broad spectral filtering width $\Gamma$ local interaction
of the pulses can be too weak against the background noise. Therefore
a laser with relatively narrow spectral filter bandwidth generating
sufficiently broad pulses is required for experimental observation
of the local TCS interaction leading to HML regimes.

To conclude, the interaction equations governing the time evolution
of the coordinates and phases of locally and non-locally interacting
TCSs in a ML NOLM-NALM laser have been derived. It has been shown
that in the case of non-local TCSs interaction due gain depletion
and slow recovery apart from the usual repulsion leading to the development
of HML regime an attractive interaction is also possible. It has been
demonstrated numerically that when the distance between two attractively
interacting TCS becomes sufficiently small an incoherent closely packed
TCS bound state similar to that observed experimentally in \cite{kokhanovskiy2020single}
can be formed, This bound state having a similarity with the bound
state reported in a Haus master equation model of a ML laser with
a slow saturable absorber \cite{soto1999multisoliton} is weakly oscillating
and affected by both the non-local and local interaction. Local interaction,
unlike the non-local one, depends strongly on the phase difference
of the interacting TCSs. In the case of two locally interacting TCSs
in a laser with relatively narrow spectral filter in-phase or anti-phase
HML regime can develop depending on the value of the pump parameter.
Moreover, a bistability between stationary and breathing HML regimes
is also possible. Note that the approach presented here is very general
and applicable to describe TCS interaction in a wide class of ML lasers,
which can be modeled by DDEs of the form (\ref{eq:general_form}).
\begin{acknowledgments}
The support by the Deutsche Forschungsgemeinschaft (DFG-RSF project
No. 445430311) is gratefully acknowledged. The author is thankful
to D. Turaev, S. Yanchuk, and D. Rachinskii for useful discussions. 
\end{acknowledgments}

\appendix

\section{Derivation of Eqs. (\ref{eq:int2_tau}) and (\ref{eq:int2_phi}).}

Linear operator ${\cal L}$ describing the stability of the TCS solution
of Eq. (\ref{eq:general_form}) and the adjoint operator ${\cal L^{\dagger}}$
are given by : 
\begin{gather}
\mathcal{L}\mathbf{v}=-\partial_{t}\mathbf{v}+\mathcal{B}(t)\mathbf{\boldsymbol{v}}+\mathcal{C}(t-\tau)\mathbf{v}_{\tau},\label{eq:L}\\
\mathcal{L}^{\dagger}\mathbf{w}=\partial_{t}\mathbf{w}+\mathbf{w}\mathcal{B}(t)+\mathbf{w}(t+\tau)\mathcal{C}(t),\label{eq:L_adjoint}
\end{gather}
where ${\bf v}$ (${\bf w}$) is a column (row) vector and the matrices
${\cal B}$ and ${\cal C}$ are obtained by linearization of ${\bf F}$
and ${\bf H}$ on the TCS solution ${\bf U}={\bf u}_{0}$ and $\omega=\omega_{0}$.

Using the fact that ${\bf u}_{1}$ and ${\bf u}_{2}$ in Eq. (\ref{eq:Anzatz})
are the solutions of Eq. (\ref{eq:general_form}) the quantity ${\bf P}$
defined by Eq. (\ref{eq:P}) can be rewritten in the form ${\bf P}={\bf Q}+{\bf S}\left(t-\tau\right),$where
${\bf Q}={\bf F}\left(\mathbf{u}_{\Sigma},\omega_{0}\right)-\sum_{k=1}^{2}{\bf F}\left(\mathbf{u}_{k},\omega_{0}\right)$
and ${\bf S}={\bf H}\left(\mathbf{u}_{\Sigma},\omega_{0}\right)-\sum_{k=1}^{2}{\bf H}\left(\mathbf{u}_{k},\omega_{0}\right)$.

Let us for simplicity consider the RHS of the interaction equations
for the TCS with the index $2$. The RHS of the interaction equation
for the first TCS can be obtained in a similar way. Furthermore, to
avoid complicating the notation we omit the superscripts $\tau$ and
$\phi$ in the adjoint neutral modes. The right hand sides of the
interaction equations are then given by 
\begin{gather}
\left\langle {\bf w}_{2}{\bf P}\right\rangle =\left\langle \mathbf{w}_{2}\left[{\bf Q}+{\bf S}\left(t-\tau\right)\right]\right\rangle =\left\langle \mathbf{w}_{2}{\bf Q}\right\rangle +\left\langle \mathbf{w}_{2}\left(t+\tau\right){\bf S}\right\rangle \nonumber \\
=\sum_{k=1}^{2}\left[\left\langle \mathbf{w}_{2}{\bf Q}\right\rangle _{k}+\left\langle \mathbf{w}_{2}\left(t+\tau\right){\bf S}\right\rangle _{k}\right],\label{eq:RHS1}
\end{gather}
where I have used $\tau_{0}$-periodicity of the TCSs and their adjoint
neutral modes and split the integral over the interval $\left[0,\tau_{0}\right]$
into two parts $\left\langle \cdot\right\rangle =\left\langle \cdot\right\rangle _{1}+\left\langle \cdot\right\rangle _{2}$
with $\left\langle \cdot\right\rangle _{1}\equiv\int_{0}^{\tau_{0}/2}\cdot dt$
and $\left\langle \cdot\right\rangle _{2}\equiv\int_{\tau_{0}/2}^{\tau_{0}}\cdot dt$.
Here without the loss of generality it is assumed that the coordinate
origin is located in the middle point between two TCSs, see Fig. \ref{fig:scheme}.
Since ${\bf u}_{1}$ is small on the integration interval $\left[0,\tau_{0}/2\right]$,
we obtain ${\bf Q}\approx\left(\mathcal{B}_{2}-\mathcal{B}_{0}\right)\mathbf{u}_{1}$
and ${\bf S}\approx\left(\mathcal{C}_{2}-\mathcal{C}_{0}\right)\mathbf{u}_{1}$
on this interval. Here the linearization matrices $\mathcal{B}_{2}\left(t\right)$
and $\mathcal{C}_{2}\left(t\right)$ are similar to $\mathcal{B}\left(t\right)$
and $\mathcal{C}\left(t\right)$ in Eqs. (\ref{eq:L}) and (\ref{eq:L_adjoint}),
but evaluated on ${\bf u}_{2}$ instead of ${\bf u}_{0}$, while $\mathcal{B}_{0}$
and $\mathcal{C}_{0}$ are the linearizations of ${\bf F}$ and ${\bf H}$
on the trivial solution ${\bf U}=0$. Similarly, on the interval $\left[\tau_{0}/2,\tau_{0}\right]$
the second TCS ${\bf u}_{2}$ is small and one can write ${\bf Q}\approx\left(\mathcal{B}_{1}-\mathcal{B}_{0}\right)\mathbf{u}_{2}$
and ${\bf S}\approx\left(\mathcal{C}_{1}-\mathcal{C}_{0}\right)\mathbf{u}_{2}$,
where $\mathcal{B}_{1}\left(t\right)$ and $\mathcal{C}_{1}\left(t\right)$
are the matrices $\mathcal{B}\left(t\right)$ and $\mathcal{C}\left(t\right)$
evaluated on ${\bf u}_{1}$ instead of ${\bf u}_{0}$. Substituting
these approximate relations into (\ref{eq:RHS1}) and neglecting the
second order terms containing ${\bf w}_{2}$ and ${\bf u}_{2}$ on
the interval $\left[\tau_{0}/2,\tau_{0}\right]$ yields
\begin{equation}
\left\langle {\bf w}_{2}{\bf P}\right\rangle \approx\left\langle \left[\mathbf{w}_{2}\left(\mathcal{B}_{2}-\mathcal{B}_{0}\right)+\mathbf{w}_{2}(t+\tau)\left(\mathcal{C}_{2}-\mathcal{C}_{0}\right)\right]\mathbf{u}_{1}\right\rangle _{1}.\label{eq:RHS2}
\end{equation}
Using the relation ${\cal L}_{2}{\bf w}_{2}=\partial_{t}\mathbf{w}_{2}+\mathbf{w}_{2}\mathcal{B}_{2}+\mathbf{w}_{2}(t+\tau)\mathcal{C}_{2}=0$,
where ${\cal L}_{2}$ is the linear operator ${\cal L}$ evaluated
on the solution ${\bf u}_{2}$ instead of ${\bf u}_{0}$, one gets
\[
\left\langle {\bf w}_{2}{\bf P}\right\rangle \approx-\left\langle \left[\partial_{t}\mathbf{w}_{2}+\mathbf{w}_{2}\mathcal{B}_{0}+\mathbf{w}_{2}(t+T)\mathcal{C}_{0}\right]\mathbf{u}_{1}\right\rangle _{1}.
\]
Finally using the relation ${\cal L}_{0}{\bf u}_{1}=-\partial_{t}\mathbf{u}_{1}+\mathcal{B}_{0}\mathbf{u}_{1}+\mathcal{C}_{0}{\bf u}_{1}(t-\tau)\approx0$
on the interval $\left[0,\tau_{0}/2\right]$, where $\mathbf{u}_{1}$
is small, one obtains 
\begin{gather}
\left\langle {\bf w}_{2}{\bf P}\right\rangle \approx-\left\langle \left[\left(\partial_{t}\mathbf{w}_{2}\right){\bf u}_{1}+\mathbf{w}_{2}\partial_{t}{\bf u}_{1}\right.\right.\nonumber \\
\left.\left.+\mathbf{w}_{2}(t+T)\mathcal{C}_{0}{\bf u}_{1}-\mathbf{w}_{2}\mathcal{C}_{0}{\bf u}_{1}\left(t-\tau\right)\right]\mathbf{u}_{1}\right\rangle _{1}=\nonumber \\
\mathbf{w}_{2}\left(0\right){\bf u}_{1}\left(0\right)-\mathbf{w}_{2}\left(\tau_{0}/2\right){\bf u}_{1}\left(\tau_{0}/2\right).\label{eq:RHS3}
\end{gather}
Here the relation ${\cal C}_{0}=0$, which is valid for the symmetric
splitter, has been used. Similarly to (\ref{eq:RHS3}), one obtains
$\left\langle {\bf w}_{1}{\bf P}\right\rangle \approx\mathbf{w}_{1}\left(\tau_{0}/2\right){\bf u}_{2}\left(\tau_{0}/2\right)-\mathbf{w}_{1}\left(\tau_{0}\right){\bf u}_{2}\left(\tau_{0}\right)=-\mathbf{w}_{1}\left(0\right){\bf u}_{2}\left(0\right)+\mathbf{w}_{1}\left(\tau_{0}/2\right){\bf u}_{2}\left(\tau_{0}/2\right).$
 \bibliographystyle{apsrev4-1}

\begin{thebibliography}{35}%
\makeatletter
\providecommand \@ifxundefined [1]{%
 \@ifx{#1\undefined}
}%
\providecommand \@ifnum [1]{%
 \ifnum #1\expandafter \@firstoftwo
 \else \expandafter \@secondoftwo
 \fi
}%
\providecommand \@ifx [1]{%
 \ifx #1\expandafter \@firstoftwo
 \else \expandafter \@secondoftwo
 \fi
}%
\providecommand \natexlab [1]{#1}%
\providecommand \enquote  [1]{``#1''}%
\providecommand \bibnamefont  [1]{#1}%
\providecommand \bibfnamefont [1]{#1}%
\providecommand \citenamefont [1]{#1}%
\providecommand \href@noop [0]{\@secondoftwo}%
\providecommand \href [0]{\begingroup \@sanitize@url \@href}%
\providecommand \@href[1]{\@@startlink{#1}\@@href}%
\providecommand \@@href[1]{\endgroup#1\@@endlink}%
\providecommand \@sanitize@url [0]{\catcode `\\12\catcode `\$12\catcode
  `\&12\catcode `\#12\catcode `\^12\catcode `\_12\catcode `\%12\relax}%
\providecommand \@@startlink[1]{}%
\providecommand \@@endlink[0]{}%
\providecommand \url  [0]{\begingroup\@sanitize@url \@url }%
\providecommand \@url [1]{\endgroup\@href {#1}{\urlprefix }}%
\providecommand \urlprefix  [0]{URL }%
\providecommand \Eprint [0]{\href }%
\providecommand \doibase [0]{http://dx.doi.org/}%
\providecommand \selectlanguage [0]{\@gobble}%
\providecommand \bibinfo  [0]{\@secondoftwo}%
\providecommand \bibfield  [0]{\@secondoftwo}%
\providecommand \translation [1]{[#1]}%
\providecommand \BibitemOpen [0]{}%
\providecommand \bibitemStop [0]{}%
\providecommand \bibitemNoStop [0]{.\EOS\space}%
\providecommand \EOS [0]{\spacefactor3000\relax}%
\providecommand \BibitemShut  [1]{\csname bibitem#1\endcsname}%
\let\auto@bib@innerbib\@empty
\bibitem [{\citenamefont {Leo}\ \emph {et~al.}(2010)\citenamefont {Leo},
  \citenamefont {Coen}, \citenamefont {Kockaert}, \citenamefont {Gorza},
  \citenamefont {Emplit},\ and\ \citenamefont {Haelterman}}]{leo2010temporal}%
  \BibitemOpen
  \bibfield  {author} {\bibinfo {author} {\bibfnamefont {F.}~\bibnamefont
  {Leo}}, \bibinfo {author} {\bibfnamefont {S.}~\bibnamefont {Coen}}, \bibinfo
  {author} {\bibfnamefont {P.}~\bibnamefont {Kockaert}}, \bibinfo {author}
  {\bibfnamefont {S.-P.}\ \bibnamefont {Gorza}}, \bibinfo {author}
  {\bibfnamefont {P.}~\bibnamefont {Emplit}}, \ and\ \bibinfo {author}
  {\bibfnamefont {M.}~\bibnamefont {Haelterman}},\ }\href@noop {} {\bibfield
  {journal} {\bibinfo  {journal} {Nature Photonics}\ }\textbf {\bibinfo
  {volume} {4}},\ \bibinfo {pages} {471} (\bibinfo {year} {2010})}\BibitemShut
  {NoStop}%
\bibitem [{\citenamefont {Herr}\ \emph {et~al.}(2014)\citenamefont {Herr},
  \citenamefont {Brasch}, \citenamefont {Jost}, \citenamefont {Wang},
  \citenamefont {Kondratiev}, \citenamefont {Gorodetsky},\ and\ \citenamefont
  {Kippenberg}}]{herr2014temporal}%
  \BibitemOpen
  \bibfield  {author} {\bibinfo {author} {\bibfnamefont {T.}~\bibnamefont
  {Herr}}, \bibinfo {author} {\bibfnamefont {V.}~\bibnamefont {Brasch}},
  \bibinfo {author} {\bibfnamefont {J.~D.}\ \bibnamefont {Jost}}, \bibinfo
  {author} {\bibfnamefont {C.~Y.}\ \bibnamefont {Wang}}, \bibinfo {author}
  {\bibfnamefont {N.~M.}\ \bibnamefont {Kondratiev}}, \bibinfo {author}
  {\bibfnamefont {M.~L.}\ \bibnamefont {Gorodetsky}}, \ and\ \bibinfo {author}
  {\bibfnamefont {T.~J.}\ \bibnamefont {Kippenberg}},\ }\href@noop {}
  {\bibfield  {journal} {\bibinfo  {journal} {Nature Photonics}\ }\textbf
  {\bibinfo {volume} {8}},\ \bibinfo {pages} {145} (\bibinfo {year}
  {2014})}\BibitemShut {NoStop}%
\bibitem [{\citenamefont {Kippenberg}\ \emph {et~al.}(2018)\citenamefont
  {Kippenberg}, \citenamefont {Gaeta}, \citenamefont {Lipson},\ and\
  \citenamefont {Gorodetsky}}]{kippenberg2018dissipative}%
  \BibitemOpen
  \bibfield  {author} {\bibinfo {author} {\bibfnamefont {T.~J.}\ \bibnamefont
  {Kippenberg}}, \bibinfo {author} {\bibfnamefont {A.~L.}\ \bibnamefont
  {Gaeta}}, \bibinfo {author} {\bibfnamefont {M.}~\bibnamefont {Lipson}}, \
  and\ \bibinfo {author} {\bibfnamefont {M.~L.}\ \bibnamefont {Gorodetsky}},\
  }\href@noop {} {\bibfield  {journal} {\bibinfo  {journal} {Science}\ }\textbf
  {\bibinfo {volume} {361}} (\bibinfo {year} {2018})}\BibitemShut {NoStop}%
\bibitem [{\citenamefont {Grelu}\ and\ \citenamefont
  {Akhmediev}(2012)}]{grelu2012dissipative}%
  \BibitemOpen
  \bibfield  {author} {\bibinfo {author} {\bibfnamefont {P.}~\bibnamefont
  {Grelu}}\ and\ \bibinfo {author} {\bibfnamefont {N.}~\bibnamefont
  {Akhmediev}},\ }\href@noop {} {\bibfield  {journal} {\bibinfo  {journal}
  {Nature photonics}\ }\textbf {\bibinfo {volume} {6}},\ \bibinfo {pages} {84}
  (\bibinfo {year} {2012})}\BibitemShut {NoStop}%
\bibitem [{\citenamefont {Marconi}\ \emph {et~al.}(2014)\citenamefont
  {Marconi}, \citenamefont {Javaloyes}, \citenamefont {Balle},\ and\
  \citenamefont {Giudici}}]{Marconi}%
  \BibitemOpen
  \bibfield  {author} {\bibinfo {author} {\bibfnamefont {M.}~\bibnamefont
  {Marconi}}, \bibinfo {author} {\bibfnamefont {J.}~\bibnamefont {Javaloyes}},
  \bibinfo {author} {\bibfnamefont {S.}~\bibnamefont {Balle}}, \ and\ \bibinfo
  {author} {\bibfnamefont {M.}~\bibnamefont {Giudici}},\ }\href {\doibase
  10.1103/PhysRevLett.112.223901} {\bibfield  {journal} {\bibinfo  {journal}
  {Phys. Rev. Lett.}\ }\textbf {\bibinfo {volume} {112}},\ \bibinfo {pages}
  {223901} (\bibinfo {year} {2014})}\BibitemShut {NoStop}%
\bibitem [{\citenamefont {Soto-Crespo}\ and\ \citenamefont
  {Akhmediev}(1999)}]{soto1999multisoliton}%
  \BibitemOpen
  \bibfield  {author} {\bibinfo {author} {\bibfnamefont {J.~M.}\ \bibnamefont
  {Soto-Crespo}}\ and\ \bibinfo {author} {\bibfnamefont {N.}~\bibnamefont
  {Akhmediev}},\ }\href@noop {} {\bibfield  {journal} {\bibinfo  {journal}
  {JOSA B}\ }\textbf {\bibinfo {volume} {16}},\ \bibinfo {pages} {674}
  (\bibinfo {year} {1999})}\BibitemShut {NoStop}%
\bibitem [{\citenamefont {Grelu}\ and\ \citenamefont
  {Soto-Crespo}(2004)}]{grelu2004multisoliton}%
  \BibitemOpen
  \bibfield  {author} {\bibinfo {author} {\bibfnamefont {P.}~\bibnamefont
  {Grelu}}\ and\ \bibinfo {author} {\bibfnamefont {J.}~\bibnamefont
  {Soto-Crespo}},\ }\href@noop {} {\bibfield  {journal} {\bibinfo  {journal}
  {Journal of Optics B: Quantum and Semiclassical Optics}\ }\textbf {\bibinfo
  {volume} {6}},\ \bibinfo {pages} {S271} (\bibinfo {year} {2004})}\BibitemShut
  {NoStop}%
\bibitem [{\citenamefont {Nizette}\ and\ \citenamefont
  {Vladimirov}(2021)}]{nizette2021generalized}%
  \BibitemOpen
  \bibfield  {author} {\bibinfo {author} {\bibfnamefont {M.}~\bibnamefont
  {Nizette}}\ and\ \bibinfo {author} {\bibfnamefont {A.~G.}\ \bibnamefont
  {Vladimirov}},\ }\href {\doibase 10.1103/PhysRevE.104.014215} {\bibfield
  {journal} {\bibinfo  {journal} {Phys. Rev. E}\ }\textbf {\bibinfo {volume}
  {104}},\ \bibinfo {pages} {014215} (\bibinfo {year} {2021})}\BibitemShut
  {NoStop}%
\bibitem [{\citenamefont {Kutz}\ \emph {et~al.}(1998)\citenamefont {Kutz},
  \citenamefont {Collings}, \citenamefont {Bergman},\ and\ \citenamefont
  {Knox}}]{kutz1998stabilized}%
  \BibitemOpen
  \bibfield  {author} {\bibinfo {author} {\bibfnamefont {J.~N.}\ \bibnamefont
  {Kutz}}, \bibinfo {author} {\bibfnamefont {B.}~\bibnamefont {Collings}},
  \bibinfo {author} {\bibfnamefont {K.}~\bibnamefont {Bergman}}, \ and\
  \bibinfo {author} {\bibfnamefont {W.}~\bibnamefont {Knox}},\ }\href@noop {}
  {\bibfield  {journal} {\bibinfo  {journal} {IEEE journal of quantum
  electronics}\ }\textbf {\bibinfo {volume} {34}},\ \bibinfo {pages} {1749}
  (\bibinfo {year} {1998})}\BibitemShut {NoStop}%
\bibitem [{\citenamefont {Akhmediev}\ \emph {et~al.}(2001)\citenamefont
  {Akhmediev}, \citenamefont {Rodrigues},\ and\ \citenamefont
  {Town}}]{akhmediev2001interaction}%
  \BibitemOpen
  \bibfield  {author} {\bibinfo {author} {\bibfnamefont {N.}~\bibnamefont
  {Akhmediev}}, \bibinfo {author} {\bibfnamefont {A.~S.}\ \bibnamefont
  {Rodrigues}}, \ and\ \bibinfo {author} {\bibfnamefont {G.~E.}\ \bibnamefont
  {Town}},\ }\href@noop {} {\bibfield  {journal} {\bibinfo  {journal} {Optics
  communications}\ }\textbf {\bibinfo {volume} {187}},\ \bibinfo {pages} {419}
  (\bibinfo {year} {2001})}\BibitemShut {NoStop}%
\bibitem [{\citenamefont {Liu}(2011)}]{liu2011interaction}%
  \BibitemOpen
  \bibfield  {author} {\bibinfo {author} {\bibfnamefont {X.}~\bibnamefont
  {Liu}},\ }\href@noop {} {\bibfield  {journal} {\bibinfo  {journal} {Phys.
  Rev. A}\ }\textbf {\bibinfo {volume} {84}},\ \bibinfo {pages} {053828}
  (\bibinfo {year} {2011})}\BibitemShut {NoStop}%
\bibitem [{\citenamefont {Peng}\ \emph {et~al.}(2019)\citenamefont {Peng},
  \citenamefont {Boscolo}, \citenamefont {Zhao},\ and\ \citenamefont
  {Zeng}}]{peng2019breathing}%
  \BibitemOpen
  \bibfield  {author} {\bibinfo {author} {\bibfnamefont {J.}~\bibnamefont
  {Peng}}, \bibinfo {author} {\bibfnamefont {S.}~\bibnamefont {Boscolo}},
  \bibinfo {author} {\bibfnamefont {Z.}~\bibnamefont {Zhao}}, \ and\ \bibinfo
  {author} {\bibfnamefont {H.}~\bibnamefont {Zeng}},\ }\href@noop {} {\bibfield
   {journal} {\bibinfo  {journal} {Science advances}\ }\textbf {\bibinfo
  {volume} {5}},\ \bibinfo {pages} {eaax1110} (\bibinfo {year}
  {2019})}\BibitemShut {NoStop}%
\bibitem [{\citenamefont {Ablowitz}\ \emph {et~al.}(2009)\citenamefont
  {Ablowitz}, \citenamefont {Horikis},\ and\ \citenamefont
  {Nixon}}]{ablowitz2009soliton}%
  \BibitemOpen
  \bibfield  {author} {\bibinfo {author} {\bibfnamefont {M.~J.}\ \bibnamefont
  {Ablowitz}}, \bibinfo {author} {\bibfnamefont {T.~P.}\ \bibnamefont
  {Horikis}}, \ and\ \bibinfo {author} {\bibfnamefont {S.~D.}\ \bibnamefont
  {Nixon}},\ }\href@noop {} {\bibfield  {journal} {\bibinfo  {journal} {Optics
  communications}\ }\textbf {\bibinfo {volume} {282}},\ \bibinfo {pages} {4127}
  (\bibinfo {year} {2009})}\BibitemShut {NoStop}%
\bibitem [{\citenamefont {Wang}\ \emph {et~al.}(2019)\citenamefont {Wang},
  \citenamefont {Nithyanandan}, \citenamefont {Coillet}, \citenamefont
  {Tchofo-Dinda},\ and\ \citenamefont {Grelu}}]{wang2019optical}%
  \BibitemOpen
  \bibfield  {author} {\bibinfo {author} {\bibfnamefont {Z.}~\bibnamefont
  {Wang}}, \bibinfo {author} {\bibfnamefont {K.}~\bibnamefont {Nithyanandan}},
  \bibinfo {author} {\bibfnamefont {A.}~\bibnamefont {Coillet}}, \bibinfo
  {author} {\bibfnamefont {P.}~\bibnamefont {Tchofo-Dinda}}, \ and\ \bibinfo
  {author} {\bibfnamefont {P.}~\bibnamefont {Grelu}},\ }\href@noop {}
  {\bibfield  {journal} {\bibinfo  {journal} {Nature communications}\ }\textbf
  {\bibinfo {volume} {10}},\ \bibinfo {pages} {1} (\bibinfo {year}
  {2019})}\BibitemShut {NoStop}%
\bibitem [{\citenamefont {Kokhanovskiy}\ \emph {et~al.}(2020)\citenamefont
  {Kokhanovskiy}, \citenamefont {Kuprikov},\ and\ \citenamefont
  {Kobtsev}}]{kokhanovskiy2020single}%
  \BibitemOpen
  \bibfield  {author} {\bibinfo {author} {\bibfnamefont {A.}~\bibnamefont
  {Kokhanovskiy}}, \bibinfo {author} {\bibfnamefont {E.}~\bibnamefont
  {Kuprikov}}, \ and\ \bibinfo {author} {\bibfnamefont {S.}~\bibnamefont
  {Kobtsev}},\ }\href@noop {} {\bibfield  {journal} {\bibinfo  {journal}
  {Optics \& Laser Technology}\ }\textbf {\bibinfo {volume} {131}},\ \bibinfo
  {pages} {106422} (\bibinfo {year} {2020})}\BibitemShut {NoStop}%
\bibitem [{\citenamefont {Puzyrev}\ \emph {et~al.}(2017)\citenamefont
  {Puzyrev}, \citenamefont {Vladimirov}, \citenamefont {Pimenov}, \citenamefont
  {Gurevich},\ and\ \citenamefont {Yanchuk}}]{puzyrev2017bound}%
  \BibitemOpen
  \bibfield  {author} {\bibinfo {author} {\bibfnamefont {D.}~\bibnamefont
  {Puzyrev}}, \bibinfo {author} {\bibfnamefont {A.}~\bibnamefont {Vladimirov}},
  \bibinfo {author} {\bibfnamefont {A.}~\bibnamefont {Pimenov}}, \bibinfo
  {author} {\bibfnamefont {S.}~\bibnamefont {Gurevich}}, \ and\ \bibinfo
  {author} {\bibfnamefont {S.}~\bibnamefont {Yanchuk}},\ }\href@noop {}
  {\bibfield  {journal} {\bibinfo  {journal} {Phys. Rev. Lett.}\ }\textbf
  {\bibinfo {volume} {119}},\ \bibinfo {pages} {163901} (\bibinfo {year}
  {2017})}\BibitemShut {NoStop}%
\bibitem [{\citenamefont {Nizette}\ \emph {et~al.}(2006)\citenamefont
  {Nizette}, \citenamefont {Rachinskii}, \citenamefont {Vladimirov},\ and\
  \citenamefont {Wolfrum}}]{nizette2006pulse}%
  \BibitemOpen
  \bibfield  {author} {\bibinfo {author} {\bibfnamefont {M.}~\bibnamefont
  {Nizette}}, \bibinfo {author} {\bibfnamefont {D.}~\bibnamefont {Rachinskii}},
  \bibinfo {author} {\bibfnamefont {A.}~\bibnamefont {Vladimirov}}, \ and\
  \bibinfo {author} {\bibfnamefont {M.}~\bibnamefont {Wolfrum}},\ }\href@noop
  {} {\bibfield  {journal} {\bibinfo  {journal} {Physica D: Nonlinear
  Phenomena}\ }\textbf {\bibinfo {volume} {218}},\ \bibinfo {pages} {95}
  (\bibinfo {year} {2006})}\BibitemShut {NoStop}%
\bibitem [{\citenamefont {Zaviyalov}\ \emph {et~al.}(2012)\citenamefont
  {Zaviyalov}, \citenamefont {Grelu},\ and\ \citenamefont
  {Lederer}}]{zaviyalov2012impact}%
  \BibitemOpen
  \bibfield  {author} {\bibinfo {author} {\bibfnamefont {A.}~\bibnamefont
  {Zaviyalov}}, \bibinfo {author} {\bibfnamefont {P.}~\bibnamefont {Grelu}}, \
  and\ \bibinfo {author} {\bibfnamefont {F.}~\bibnamefont {Lederer}},\
  }\href@noop {} {\bibfield  {journal} {\bibinfo  {journal} {Optics letters}\
  }\textbf {\bibinfo {volume} {37}},\ \bibinfo {pages} {175} (\bibinfo {year}
  {2012})}\BibitemShut {NoStop}%
\bibitem [{\citenamefont {Javaloyes}\ \emph {et~al.}(2016)\citenamefont
  {Javaloyes}, \citenamefont {Camelin}, \citenamefont {Marconi},\ and\
  \citenamefont {Giudici}}]{javaloyes2016dynamics}%
  \BibitemOpen
  \bibfield  {author} {\bibinfo {author} {\bibfnamefont {J.}~\bibnamefont
  {Javaloyes}}, \bibinfo {author} {\bibfnamefont {P.}~\bibnamefont {Camelin}},
  \bibinfo {author} {\bibfnamefont {M.}~\bibnamefont {Marconi}}, \ and\
  \bibinfo {author} {\bibfnamefont {M.}~\bibnamefont {Giudici}},\ }\href@noop
  {} {\bibfield  {journal} {\bibinfo  {journal} {Phys. Rev. Lett.}\ }\textbf
  {\bibinfo {volume} {116}},\ \bibinfo {pages} {133901} (\bibinfo {year}
  {2016})}\BibitemShut {NoStop}%
\bibitem [{\citenamefont {Camelin}\ \emph {et~al.}(2016)\citenamefont
  {Camelin}, \citenamefont {Javaloyes}, \citenamefont {Marconi},\ and\
  \citenamefont {Giudici}}]{camelin2016electrical}%
  \BibitemOpen
  \bibfield  {author} {\bibinfo {author} {\bibfnamefont {P.}~\bibnamefont
  {Camelin}}, \bibinfo {author} {\bibfnamefont {J.}~\bibnamefont {Javaloyes}},
  \bibinfo {author} {\bibfnamefont {M.}~\bibnamefont {Marconi}}, \ and\
  \bibinfo {author} {\bibfnamefont {M.}~\bibnamefont {Giudici}},\ }\href@noop
  {} {\bibfield  {journal} {\bibinfo  {journal} {Phys. Rev. A}\ }\textbf
  {\bibinfo {volume} {94}},\ \bibinfo {pages} {063854} (\bibinfo {year}
  {2016})}\BibitemShut {NoStop}%
\bibitem [{\citenamefont {Vladimirov}\ \emph {et~al.}(2019)\citenamefont
  {Vladimirov}, \citenamefont {Kovalev}, \citenamefont {Viktorov},
  \citenamefont {Rebrova},\ and\ \citenamefont
  {Huyet}}]{vladimirov2019dynamics}%
  \BibitemOpen
  \bibfield  {author} {\bibinfo {author} {\bibfnamefont {A.~G.}\ \bibnamefont
  {Vladimirov}}, \bibinfo {author} {\bibfnamefont {A.~V.}\ \bibnamefont
  {Kovalev}}, \bibinfo {author} {\bibfnamefont {E.~A.}\ \bibnamefont
  {Viktorov}}, \bibinfo {author} {\bibfnamefont {N.}~\bibnamefont {Rebrova}}, \
  and\ \bibinfo {author} {\bibfnamefont {G.}~\bibnamefont {Huyet}},\
  }\href@noop {} {\bibfield  {journal} {\bibinfo  {journal} {Phys. Rev. E}\
  }\textbf {\bibinfo {volume} {100}},\ \bibinfo {pages} {012216} (\bibinfo
  {year} {2019})}\BibitemShut {NoStop}%
\bibitem [{\citenamefont {Vladimirov}\ and\ \citenamefont
  {Turaev}(2005)}]{VT05}%
  \BibitemOpen
  \bibfield  {author} {\bibinfo {author} {\bibfnamefont {A.~G.}\ \bibnamefont
  {Vladimirov}}\ and\ \bibinfo {author} {\bibfnamefont {D.}~\bibnamefont
  {Turaev}},\ }\href@noop {} {\bibfield  {journal} {\bibinfo  {journal} {Phys.
  Rev. A}\ }\textbf {\bibinfo {volume} {72}},\ \bibinfo {pages} {033808}
  (\bibinfo {year} {2005})}\BibitemShut {NoStop}%
\bibitem [{\citenamefont {Vladimirov}\ \emph {et~al.}(2004)\citenamefont
  {Vladimirov}, \citenamefont {Turaev},\ and\ \citenamefont {Kozyreff}}]{VTK}%
  \BibitemOpen
  \bibfield  {author} {\bibinfo {author} {\bibfnamefont {A.~G.}\ \bibnamefont
  {Vladimirov}}, \bibinfo {author} {\bibfnamefont {D.}~\bibnamefont {Turaev}},
  \ and\ \bibinfo {author} {\bibfnamefont {G.}~\bibnamefont {Kozyreff}},\
  }\href@noop {} {\bibfield  {journal} {\bibinfo  {journal} {Opt. Lett.}\
  }\textbf {\bibinfo {volume} {29}},\ \bibinfo {pages} {1221} (\bibinfo {year}
  {2004})}\BibitemShut {NoStop}%
\bibitem [{\citenamefont {Vladimirov}\ and\ \citenamefont
  {Turaev}(2004)}]{VT04}%
  \BibitemOpen
  \bibfield  {author} {\bibinfo {author} {\bibfnamefont {A.~G.}\ \bibnamefont
  {Vladimirov}}\ and\ \bibinfo {author} {\bibfnamefont {D.}~\bibnamefont
  {Turaev}},\ }\href@noop {} {\bibfield  {journal} {\bibinfo  {journal}
  {Radiophys. \& Quant. Electron.}\ }\textbf {\bibinfo {volume} {47}},\
  \bibinfo {pages} {857} (\bibinfo {year} {2004})}\BibitemShut {NoStop}%
\bibitem [{\citenamefont {Vladimirov}\ \emph
  {et~al.}(2021{\natexlab{a}})\citenamefont {Vladimirov}, \citenamefont
  {Suchkov}, \citenamefont {Huyet},\ and\ \citenamefont
  {Turitsyn}}]{vladimirov2021delay}%
  \BibitemOpen
  \bibfield  {author} {\bibinfo {author} {\bibfnamefont {A.~G.}\ \bibnamefont
  {Vladimirov}}, \bibinfo {author} {\bibfnamefont {S.}~\bibnamefont {Suchkov}},
  \bibinfo {author} {\bibfnamefont {G.}~\bibnamefont {Huyet}}, \ and\ \bibinfo
  {author} {\bibfnamefont {S.~K.}\ \bibnamefont {Turitsyn}},\ }\href@noop {}
  {\bibfield  {journal} {\bibinfo  {journal} {Phys. Rev. A}\ }\textbf {\bibinfo
  {volume} {104}},\ \bibinfo {pages} {033525} (\bibinfo {year}
  {2021}{\natexlab{a}})}\BibitemShut {NoStop}%
\bibitem [{\citenamefont {Doran}\ and\ \citenamefont
  {Wood}(1988)}]{doran1988nonlinear}%
  \BibitemOpen
  \bibfield  {author} {\bibinfo {author} {\bibfnamefont {N.}~\bibnamefont
  {Doran}}\ and\ \bibinfo {author} {\bibfnamefont {D.}~\bibnamefont {Wood}},\
  }\href@noop {} {\bibfield  {journal} {\bibinfo  {journal} {Opt. Lett.}\
  }\textbf {\bibinfo {volume} {13}},\ \bibinfo {pages} {56} (\bibinfo {year}
  {1988})}\BibitemShut {NoStop}%
\bibitem [{\citenamefont {Lai}\ \emph {et~al.}(2005)\citenamefont {Lai},
  \citenamefont {Shum},\ and\ \citenamefont {Binh}}]{lai2005nolm}%
  \BibitemOpen
  \bibfield  {author} {\bibinfo {author} {\bibfnamefont {W.~J.}\ \bibnamefont
  {Lai}}, \bibinfo {author} {\bibfnamefont {P.}~\bibnamefont {Shum}}, \ and\
  \bibinfo {author} {\bibfnamefont {L.}~\bibnamefont {Binh}},\ }\href@noop {}
  {\bibfield  {journal} {\bibinfo  {journal} {IEEE journal of quantum
  electronics}\ }\textbf {\bibinfo {volume} {41}},\ \bibinfo {pages} {986}
  (\bibinfo {year} {2005})}\BibitemShut {NoStop}%
\bibitem [{\citenamefont {Fermann}\ \emph {et~al.}(1990)\citenamefont
  {Fermann}, \citenamefont {Haberl}, \citenamefont {Hofer},\ and\ \citenamefont
  {Hochreiter}}]{fermann1990nonlinear}%
  \BibitemOpen
  \bibfield  {author} {\bibinfo {author} {\bibfnamefont {M.~E.}\ \bibnamefont
  {Fermann}}, \bibinfo {author} {\bibfnamefont {F.}~\bibnamefont {Haberl}},
  \bibinfo {author} {\bibfnamefont {M.}~\bibnamefont {Hofer}}, \ and\ \bibinfo
  {author} {\bibfnamefont {H.}~\bibnamefont {Hochreiter}},\ }\href@noop {}
  {\bibfield  {journal} {\bibinfo  {journal} {Optics Letters}\ }\textbf
  {\bibinfo {volume} {15}},\ \bibinfo {pages} {752} (\bibinfo {year}
  {1990})}\BibitemShut {NoStop}%
\bibitem [{\citenamefont {Pimenov}\ \emph {et~al.}(2017)\citenamefont
  {Pimenov}, \citenamefont {Slepneva}, \citenamefont {Huyet},\ and\
  \citenamefont {Vladimirov}}]{pimenovprl}%
  \BibitemOpen
  \bibfield  {author} {\bibinfo {author} {\bibfnamefont {A.}~\bibnamefont
  {Pimenov}}, \bibinfo {author} {\bibfnamefont {S.}~\bibnamefont {Slepneva}},
  \bibinfo {author} {\bibfnamefont {G.}~\bibnamefont {Huyet}}, \ and\ \bibinfo
  {author} {\bibfnamefont {A.~G.}\ \bibnamefont {Vladimirov}},\ }\href
  {\doibase 10.1103/PhysRevLett.118.193901} {\bibfield  {journal} {\bibinfo
  {journal} {Phys. Rev. Lett.}\ }\textbf {\bibinfo {volume} {118}},\ \bibinfo
  {pages} {193901} (\bibinfo {year} {2017})}\BibitemShut {NoStop}%
\bibitem [{\citenamefont {Pimenov}\ \emph {et~al.}(2020)\citenamefont
  {Pimenov}, \citenamefont {Amiranashvili},\ and\ \citenamefont
  {Vladimirov}}]{pimenov2020temporal}%
  \BibitemOpen
  \bibfield  {author} {\bibinfo {author} {\bibfnamefont {A.}~\bibnamefont
  {Pimenov}}, \bibinfo {author} {\bibfnamefont {S.}~\bibnamefont
  {Amiranashvili}}, \ and\ \bibinfo {author} {\bibfnamefont {A.~G.}\
  \bibnamefont {Vladimirov}},\ }\href@noop {} {\bibfield  {journal} {\bibinfo
  {journal} {Mathematical Modelling of Natural Phenomena}\ }\textbf {\bibinfo
  {volume} {15}},\ \bibinfo {pages} {47} (\bibinfo {year} {2020})}\BibitemShut
  {NoStop}%
\bibitem [{\citenamefont {Yanchuk}\ \emph {et~al.}(2019)\citenamefont
  {Yanchuk}, \citenamefont {Ruschel}, \citenamefont {Sieber},\ and\
  \citenamefont {Wolfrum}}]{yanchuk2019temporal}%
  \BibitemOpen
  \bibfield  {author} {\bibinfo {author} {\bibfnamefont {S.}~\bibnamefont
  {Yanchuk}}, \bibinfo {author} {\bibfnamefont {S.}~\bibnamefont {Ruschel}},
  \bibinfo {author} {\bibfnamefont {J.}~\bibnamefont {Sieber}}, \ and\ \bibinfo
  {author} {\bibfnamefont {M.}~\bibnamefont {Wolfrum}},\ }\href@noop {}
  {\bibfield  {journal} {\bibinfo  {journal} {Physical review letters}\
  }\textbf {\bibinfo {volume} {123}},\ \bibinfo {pages} {053901} (\bibinfo
  {year} {2019})}\BibitemShut {NoStop}%
\bibitem [{\citenamefont {Halanay}(1966)}]{halanay1966differential}%
  \BibitemOpen
  \bibfield  {author} {\bibinfo {author} {\bibfnamefont {A.}~\bibnamefont
  {Halanay}},\ }\href@noop {} {\emph {\bibinfo {title} {Differential equations:
  Stability, oscillations, time lags}}},\ Vol.~\bibinfo {volume} {6}\ (\bibinfo
   {publisher} {Elsevier},\ \bibinfo {year} {1966})\BibitemShut {NoStop}%
\bibitem [{\citenamefont {Vladimirov}\ \emph {et~al.}(2001)\citenamefont
  {Vladimirov}, \citenamefont {Khodova},\ and\ \citenamefont
  {Rosanov}}]{vladimirov2001stable}%
  \BibitemOpen
  \bibfield  {author} {\bibinfo {author} {\bibfnamefont {A.}~\bibnamefont
  {Vladimirov}}, \bibinfo {author} {\bibfnamefont {G.}~\bibnamefont {Khodova}},
  \ and\ \bibinfo {author} {\bibfnamefont {N.}~\bibnamefont {Rosanov}},\
  }\href@noop {} {\bibfield  {journal} {\bibinfo  {journal} {Phys. Rev. E}\
  }\textbf {\bibinfo {volume} {63}},\ \bibinfo {pages} {056607} (\bibinfo
  {year} {2001})}\BibitemShut {NoStop}%
\bibitem [{\citenamefont {Vladimirov}\ \emph {et~al.}(2018)\citenamefont
  {Vladimirov}, \citenamefont {Gurevich},\ and\ \citenamefont
  {Tlidi}}]{vladimirov2018effect}%
  \BibitemOpen
  \bibfield  {author} {\bibinfo {author} {\bibfnamefont {A.~G.}\ \bibnamefont
  {Vladimirov}}, \bibinfo {author} {\bibfnamefont {S.~V.}\ \bibnamefont
  {Gurevich}}, \ and\ \bibinfo {author} {\bibfnamefont {M.}~\bibnamefont
  {Tlidi}},\ }\href@noop {} {\bibfield  {journal} {\bibinfo  {journal} {Phys.
  Rev. A}\ }\textbf {\bibinfo {volume} {97}},\ \bibinfo {pages} {013816}
  (\bibinfo {year} {2018})}\BibitemShut {NoStop}%
\bibitem [{\citenamefont {Vladimirov}\ \emph
  {et~al.}(2021{\natexlab{b}})\citenamefont {Vladimirov}, \citenamefont
  {Tlidi},\ and\ \citenamefont {Taki}}]{vladimirov2021dissipative}%
  \BibitemOpen
  \bibfield  {author} {\bibinfo {author} {\bibfnamefont {A.~G.}\ \bibnamefont
  {Vladimirov}}, \bibinfo {author} {\bibfnamefont {M.}~\bibnamefont {Tlidi}}, \
  and\ \bibinfo {author} {\bibfnamefont {M.}~\bibnamefont {Taki}},\ }\href@noop
  {} {\bibfield  {journal} {\bibinfo  {journal} {Phys. Rev. A}\ }\textbf
  {\bibinfo {volume} {103}},\ \bibinfo {pages} {063505} (\bibinfo {year}
  {2021}{\natexlab{b}})}\BibitemShut {NoStop}%
\end{thebibliography}

\end{document}